\documentclass[usenatbib,usegraphicx]{mn2e}

\usepackage{amsmath}
\usepackage{subfigure}

\title[Radio observations of HD 189733 b]{Secondary radio eclipse of the transiting planet HD~189733~b: an upper limit at 307-347 MHz}

\author[A. M. S. Smith et al.]{A. M. S. Smith$^{1}$\thanks{E-mail:
amss@st-and.ac.uk},
A. Collier Cameron$^{1}$,
J. Greaves$^{1}$,
M. Jardine$^{1}$,
\newauthor
G. Langston$^{2}$,
D. Backer$^{3}$\\
$^{1}$SUPA\thanks{Scottish Universities Physics Alliance}, School of Physics \& Astronomy, University of St. Andrews, North Haugh, 
St. Andrews, Fife, KY16 9SS, UK \\
$^{2}$National Radio Astronomy Observatory, Green Bank, WV 24944, USA\\ 
$^{3}$Astronomy Department, University of California at Berkeley, Berkeley, CA 94720, USA \\
}

\begin{document}
\maketitle

\begin{abstract}

We report the first attempt to observe the secondary eclipse of a transiting extra-solar planet at radio wavelengths.  We observed HD 189733 b with the Robert C. Byrd Green Bank Telescope of the NRAO over about 5.5 hours before, during and after secondary eclipse, at frequencies of 307 - 347 MHz. In this frequency range, we determine the 3-$\sigma$ upper limit to the flux density to be 81 mJy. The data are consistent with no eclipse or a marginal reduction in flux at the time of secondary eclipse in all subsets of our bandwidth; the strongest signal is an apparent eclipse at the 2-$\sigma$ level in the 335.2 - 339.3 MHz region. Our observed upper limit is close to theoretical predictions of the flux density of cyclotron-maser radiation from the planet.

\end{abstract}
\section{Introduction}

In the solar system, all the magnetised planets are known to emit at radio wavelengths. One of the principal mechanisms producing this is electron-cyclotron maser radiation, which produces emission at the local gyrofrequency, $f_{\mathrm{g}}$, given by

\begin{equation}
\label{eqn:gyro}
f_{\mathrm{g}}(\mathrm{MHz}) = 2.8 B(\mathrm{G}),
\end{equation}
where $B$ is the planetary magnetic field strength.

Of the solar system planets, Jupiter has the strongest radio emission, outshining the (quiescent) Sun by several orders of magnitude at frequencies of about 30 MHz. Jupiter was also the first planet discovered to emit at radio wavelengths, detected at 22 MHz by \cite{Burke}, since the other solar system planets (including Earth) emit at frequencies below the Earth's ionospheric cut-off at $\sim 5 - 10$ MHz {\citep{Gmr05}.

It has long been suggested that analogous radio emission might arise in an extra-solar planetary system, and that it may be possible to detect this radiation from Earth. Indeed,  several attempts have been made to detect radio emission from an extra-solar planetary system, beginning with observations in the 1970s and 80s before any extra-solar planets were known (see \citealt{Gmr06} for a list of observational campaigns). More recently, with the discovery of hundreds of extra-solar planets starting with 51 Peg b \citep{M&Q95}, several further attempts have been made to detect radio emission from these planets, none successfully to date.

Observations of the radio emission from Jupiter have allowed the planet's magnetic field strength to be inferred from the high-frequency cut-off (around 40 MHz). The emission with this highest frequency is generated in the region with the highest magnetic field strength (equation \ref{eqn:gyro}), close to the planet. Observations of the radio spectrum of hot Jupiters could provide similar information about the planetary magnetic field, composition and rotation.

Although Jupiter's emission is several orders of magnitude stronger than the (quiescent) Sun's at certain frequencies, the analogous emission from extra-solar systems is expected to be very faint because of the distance involved. Consequently, the planetary systems most favoured by observers tend to be those within a few pc of Earth.

A new theoretical model \citep{Jardine} predicts the
radio flux from a range of exoplanets, both those in large orbits
that are embedded in the stellar wind, and those in close orbits
that lie within the stellar magnetosphere. Reconnection between
the stellar and planetary magnetic fields provides the pool of accelerated
electrons necessary for the electron-cyclotron maser instability. This
model reproduces the observed emission from solar-system planets
and predicts that the radio flux from exoplanets in close orbits 
should scale as

\begin{equation}
S_\nu \propto \left[ \frac{N}{B_\star^{1/3}} \right]^2
\end{equation}
where $N$ is the coronal density and $B_\star$ is the surface field strength
of the parent star.

In particular, this model predicts the radio flux for HD 189733 b, a 
hot-Jupiter planet orbiting a K2 dwarf \citep{Bouchy05}. For the observed
field strength of 40 G \citep{Moutou07} and a solar coronal density, 
this gives 15mJy, at a frequency determined by the planetary magnetic field strength.  It is likely, however, that the coronal density will
be greater than the solar value, since HD189733 rotates more rapidly 
than the Sun. If we assume scalings in the range $N \propto \Omega^{0.6}$ 
to  $N \propto \Omega$ (\citealt{Unruh97}, \citealt{I&T03})
then for the rotation rate of 11.73 days appropriate for HD189733, the flux 
rises to 39 - 72 mJy respectively.

In this work, we make use of the fact that HD 189733 b transits its host star to observe the system during secondary eclipse, when the planet passes behind the star. Detecting the secondary eclipse at radio wavelengths would provide unambiguous evidence that the emission is of planetary, rather than stellar, origin. These observations (described in Section \ref{sec:obs}), conducted on a bandwidth centred at 327 MHz, are used to determine new upper limits to the radio flux from HD 189733 b (Section \ref{sec:data}). This relatively low frequency was chosen because the planet could emit at this frequency given a plausibly strong magnetic field. The significance of these upper limits and the prospects of future work are discussed in Section \ref{sec:discuss}.

\section{Observations}
\label{sec:obs}

We observed the HD189733 system for $\sim$5.7 hours, encompassing a 1.827 hour \citep{Winn07} secondary eclipse on 2007 April 21, when the star was at altitudes between 45$^\circ$ and 74$^\circ$. The observations were conducted using the Spectral Processor at PF1 of the Robert C. Byrd Green Bank Telescope (GBT) of the National Radio Astronomy Observatory (NRAO)

The dual polarisation observations span the frequency range 307 MHz to 347 MHz and consist of a series of spectra comprising 1024 frequency channels, each with an integration time of one second. Spectra were grouped into scans, of duration 120 s, and scans of the target (ON) were alternated with off-target scans (OFF) of a patch of sky whose centre is 2$^\circ$ South of the target in declination.

The reason for choosing to make spectral-line observations, despite the broad-band nature of the expected signal, is to mitigate against radio frequency interference (RFI). Much RFI is narrow-band in nature, so observing with a large number of frequency channels allows channels affected by RFI to be excluded (see Section \ref{sec:rfi}).

Because of the large beam size at this frequency, we expect there to be other radio sources within the beam. The WENSS 300 MHz survey has a sensitivity of 18 mJy, but does not extend below +30$\degr$ in declination and so does not cover our target star; it can, however, give us an idea of how many objects we can expect in a typical GBT beam (FWHM = 36'). We looked in the WENSS catalogue at ten beam pointings spaced between 20 and 21 hours in RA, at +31$\degr$ in declination. We find an average of 4.8 objects within 20' of the beam centre; these objects have a median integrated flux of 121 mJy. The Texas Survey of discrete radio sources at 365 MHz \citep{Texas} is less sensitive (90 per cent complete to 400 mJy and 80 per cent to 250 mJy), but it does cover HD189733. The Texas survey includes an object with a flux density of 550 mJy 18.8' from the centre of our ON beam, and two objects of flux density 581 mJy and 241 mJy, respectively 3.6' and 16.1' from the centre of our OFF beam.

Given that we are searching for a source exhibiting a specific, known variability\footnote{Using the period and transit duration of \cite{Winn07}, and the time of mid-eclipse from \cite{Deming06}, and associated uncertainites, we conclude that the eclipse time is known to within 200 s.}, the other radio sources in our beam are unlikely to be a problem. This is because in order to mimic the eclipse we are looking for, any variable object would not only have to vary, but would have to produce a large ($\sim$ 10 per cent or greater)  eclipse-like signal on a specific, relatively short, timescale.

Each observation was calibrated using a noise tube calibrator, but we do not correct for atmospheric opacity, since such effects are negligible in the 1 m wavelength regime. Given the beam size and pointing stability of the GBT, we do not need to correct for pointing errors. An increase in raw flux was observed in both our target and off-target beams at around 2454214.0 JD which corresponds to 8 am local time and is probably caused by an increase in radio-frequency interference.

\section{Analysis}
\label{sec:data}

\subsection{Production of lightcurves}
\label{sec:rfi}

Initially, the data were visually inspected by stacking a number of spectra into a time series. A typical set of stacked spectra are shown in Figure \ref{fig:colmap2}, using colour to indicate flux. The most noticeable features in these stacked spectra are (i) the sharp 'spikes' spanning just one or two frequency channels and (ii) an oscillation with a period of around 20 s across all frequencies. Since both these effects are observed in both the ON and OFF scans, we conclude that they are due to radio-frequency interference.

In order to determine which frequency channels suffer from intermittent 'spiky' RFI, the variance of each of the 1024 channels over the whole dataset was calculated (Figure \ref{fig:var1}). After noticing a small number of pairs of scans suffer from a high level of RFI over a broad range of frequencies, we identified and excluded a total of seven pairs of scans from further analysis. It is believed that the RFI in at least some of these scans was caused by electrical interference in the control room of the GBT. The variance in each frequency channel after the exclusion of these scans (Figure \ref{fig:var2}) shows only narrow band spikes of RFI, which is easily identified and removed. Note that the same analysis of the raw OFF scans reveals spikes of high flux density in identical frequency channels, confirming that RFI, and not flaring of the target star, is responsible. A total of 177 of the 1024 channels were in this way identified as containing RFI and excluded, resulting in a variance plot without spikes (Figure \ref{fig:var3}).

\begin{figure}
\includegraphics[angle=0,width=8.25cm]{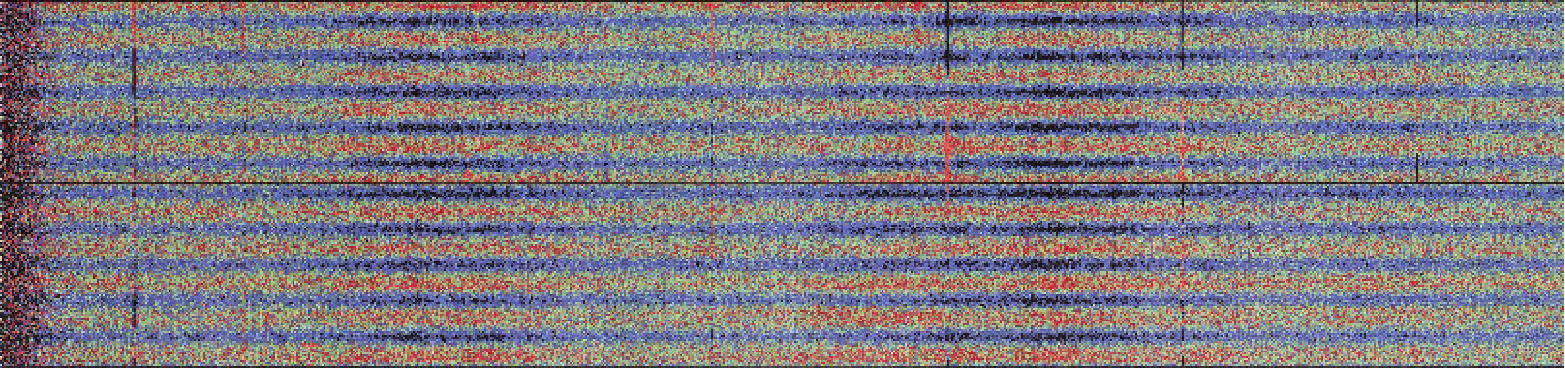}
\caption{Series of time-stacked spectra for a typical pair of scans (ON, lower half of figure; OFF, upper half), each of duration 120 s. Time increases from bottom to top; frequency from left (307 MHz) to right (347 MHz). Colours are indicative of raw flux counts, red (light) represents high flux and blue (dark) low flux.
}
\protect\label{fig:colmap2}
\end{figure}

\begin{figure}
\subfigure{\label{fig:var1}
(a)
\includegraphics[angle=270,width=8.25cm]{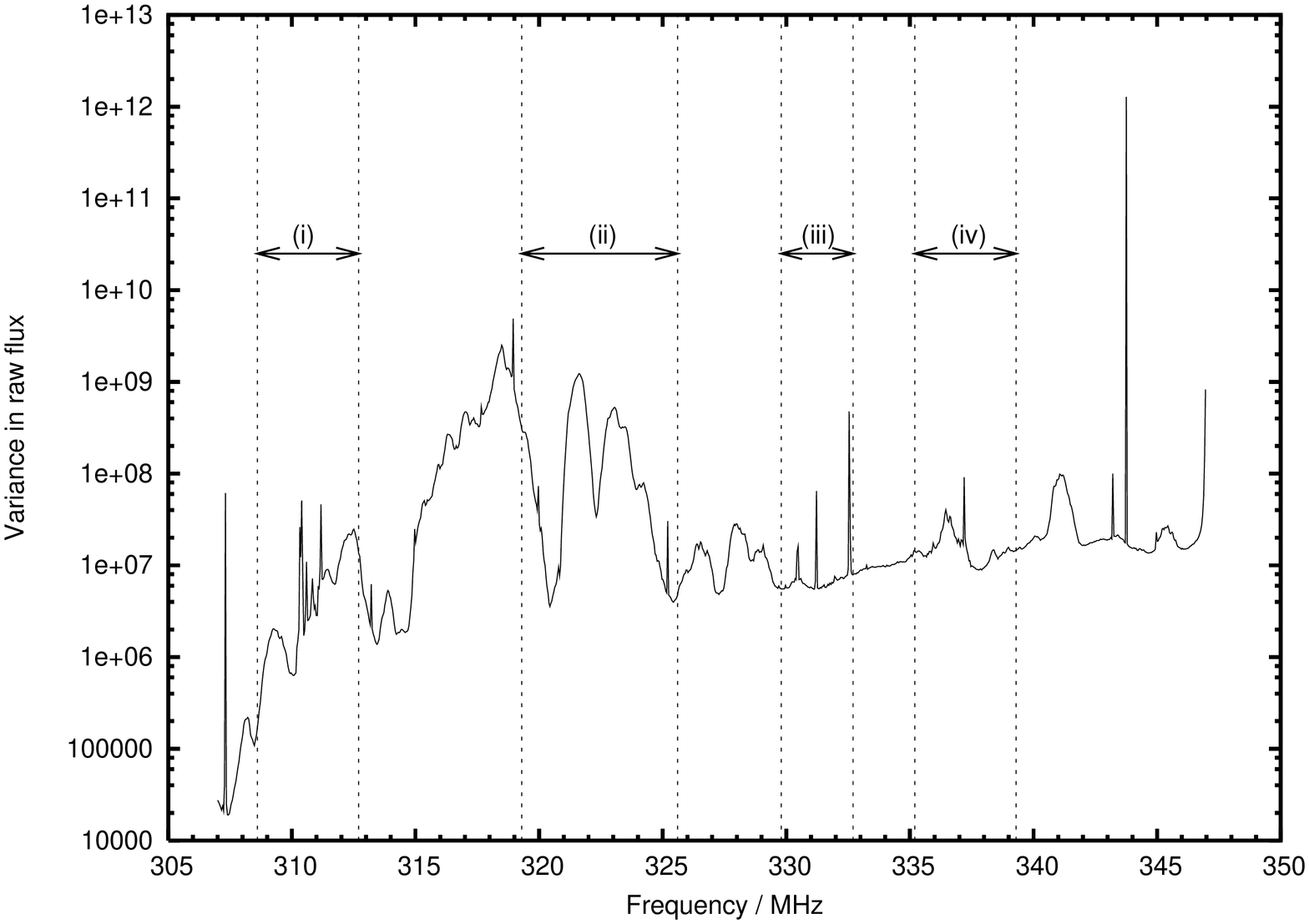}
}
(b)
\subfigure{\label{fig:var2}
\includegraphics[angle=270,width=8.25cm]{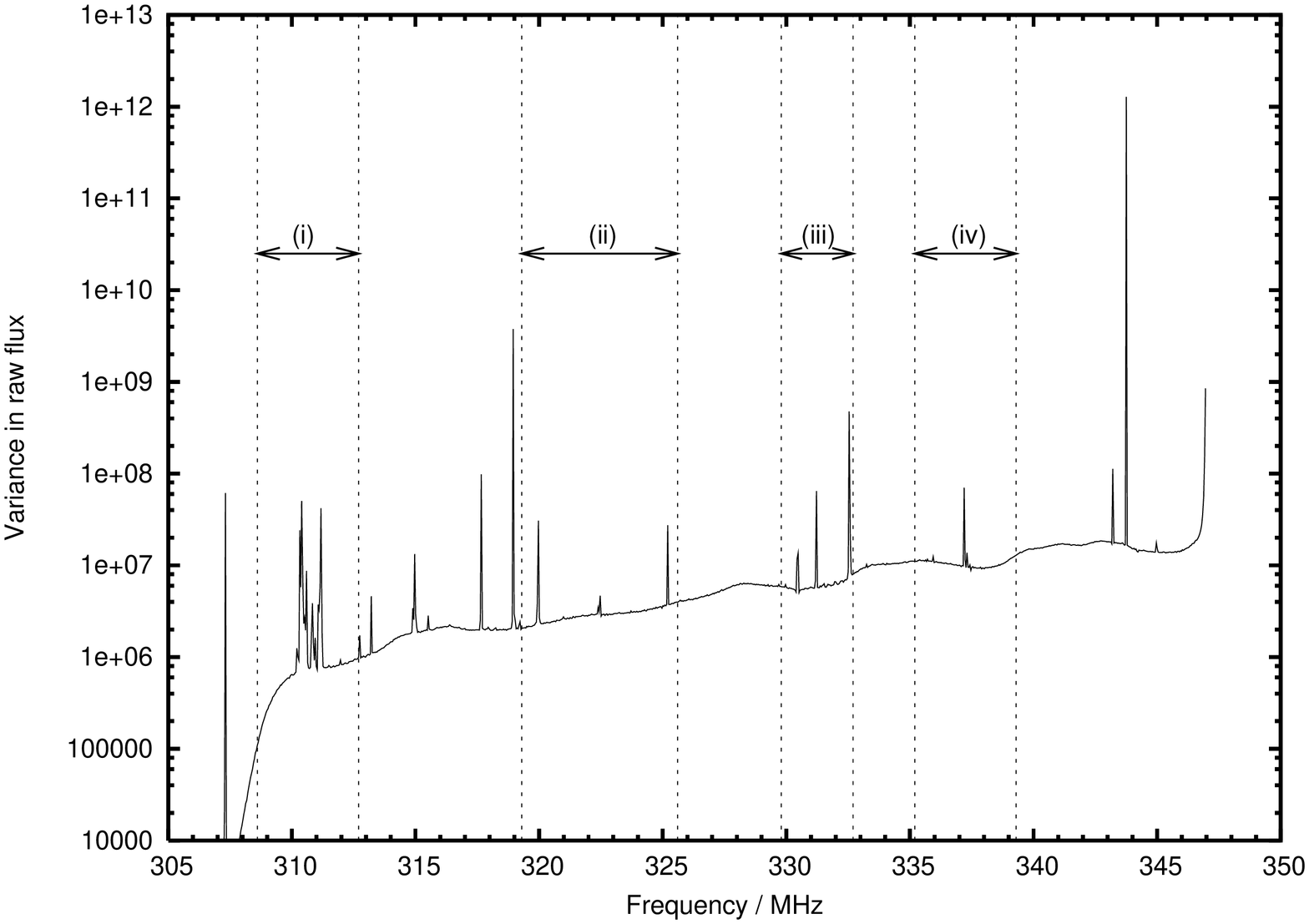}
}
(c)
\subfigure{\label{fig:var3}
\includegraphics[angle=270,width=8.25cm]{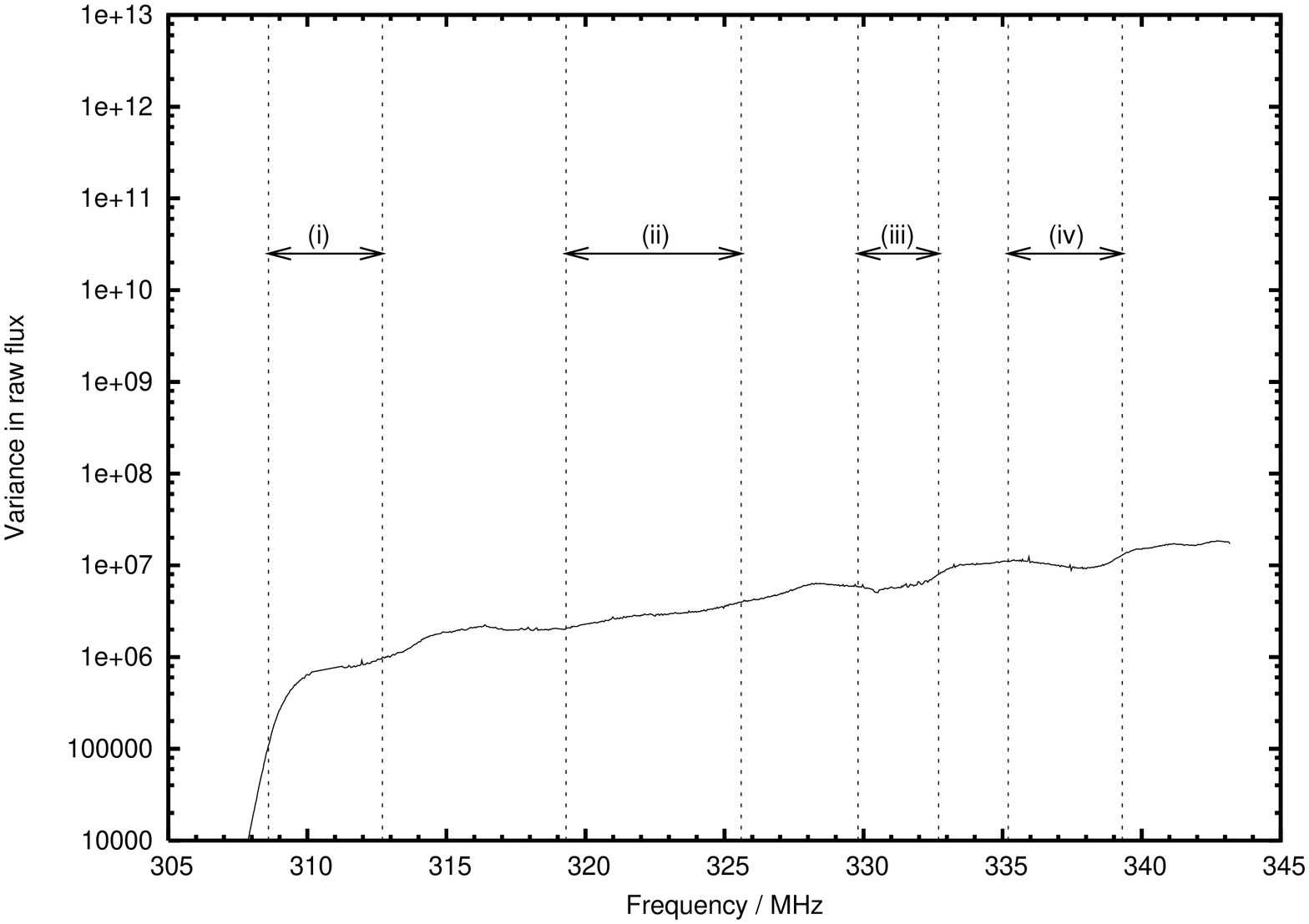}
}
\caption{Variance in the raw flux for the ON scans taken on 2007 April 23. The top panel includes all 66 ON scans and all frequency channels. The middle panel is the same as the top panel, but 7 noisy scans have been completely excluded. The bottom panel is the same as the middle panel, but with 177 noisy frequency channels removed. The four regions indicated in each panel are those parts of the bandwidth for which we generate lightcurves (see text).
}
\protect\label{fig:var}
\end{figure}

In order to determine which parts of the remaining bandpass to use to construct lightcurves and search for an eclipse signal, we calculated the standard deviation of each frequency channel of the raw ON lightcurve, which is normalised by dividing by the mean raw flux (figure \ref{fig:sdflux}). With the aid of this figure, four regions of the bandpass that show less variation than surrounding regions were identified. We number these four subsets of the bandpass (i) - (iv) and construct lightcurves from the frequency channels in each of these bandpasses, as well as one from all of the remaining 847 frequency channels.

In each case, single, uncalibrated ON and OFF lightcurves were created by taking the mean of the relevant channels at 1 s intervals. The periodic RFI identified in Figure \ref{fig:colmap2} earlier is clearly visible in plots of these ON and OFF lightcurves spanning only a few minutes (Figure \ref{fig:fit}). A rectified sine wave was used to fit this periodic modulation, which we believe to be caused by a distant radar system. The ON and OFF lightcurves for the whole bandpass, after the rectified sine wave has been subtracted, are shown in figure \ref{fig:raw}.

After subtracting the rectified sine wave from each scan (ON and OFF), the lightcurves were median-binned on the duration of a scan, and the ON scan was converted to a flux density, $S_\nu$, in Jy, according to

\begin{equation}
S_\nu = \frac{1}{\Gamma} \times \frac{(ON - OFF)}{OFF}\times T_{\mathrm{sys}},
\end{equation}
where $ON$ and $OFF$ are the raw fluxes of the ON and OFF scans respectively, $T_{\mathrm{sys}}$ is the system temperature, and $\Gamma = 2.0$ KJy$^{-1}$ is the telescope sensitivity. The resulting binned, calibrated lightcurves are shown in Figure \ref{fig:lc}. Although the residuals in our fit to the periodic RFI result in errors on individual points of magnitude $\sim$ 100 mJy, this should not affect our ability to detect a signal, given our number of data points.

\begin{figure}
\includegraphics[angle=270,width=8.25cm]{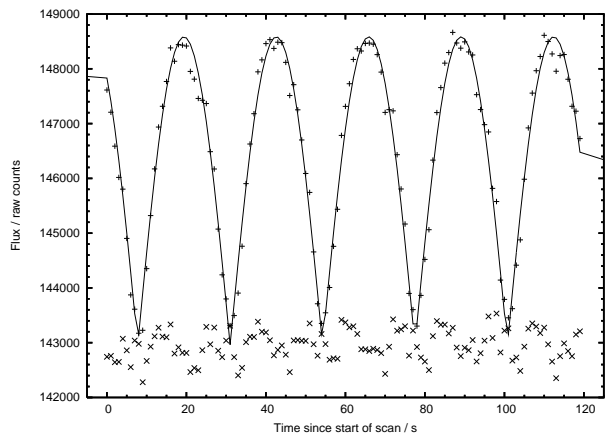}
\caption{A typical ON scan of duration 120 s showing the raw counts (averaged over all frequency channels) as a function of time (+). Also plotted are the rectified sine wave fitted to this scan (solid line) and the residuals of this fit (x).
}
\protect\label{fig:fit}
\end{figure}

\begin{figure}
\includegraphics[angle=270,width=8.25cm]{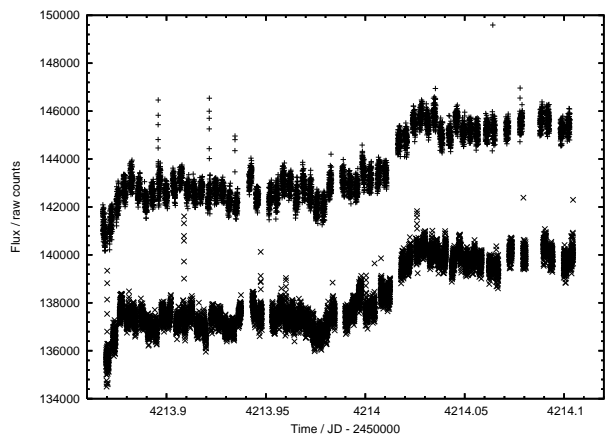}
\caption{Uncalibrated lightcurves ON (upper curve; denoted by +) and OFF (lower; x) target, after subtraction of periodic (rectified sine wave) RFI.
}
\protect\label{fig:raw}
\end{figure}

\subsection{Measuring the eclipse depth}

We use the period ($P = 2.2185733$ days), epoch of transit ($t_0 = 2453988.80336$ (HJD)), and transit duration ($t_{IV} - t_I = 1.827$ hours) determined by \cite{Winn07} and the fact that the orbit of HD 189733 b is circular to identify the 'low' points, $\ell$, observed during secondary eclipse. We then follow  the method of \cite{Cameron-etal06} to calculate the depth of transit and the associated uncertainty. The weighted mean of the lightcurve is subtracted from all lightcurve points, $x_i$, and the inverse-variance weights, $w_i$, are scaled to give $\chi ^2 = 1$ for a constant-flux model fitted to the out-of-eclipse data. The transit depth, $\delta$, and its variance, Var($\delta$) are then calculated according to,
\begin{equation}
\delta = \frac{\sum_{i \in \ell} x_i w_i \sum_i w_i}{\sum_{i \in \ell} w_i \left [ \sum_i w_i - \sum_{i \in \ell} w_i \right ]}
\end{equation}
and
\begin{equation}
\mathrm{Var}(\delta) = \frac{\sum_i w_i}{\sum_{i \in \ell} w_i \left [ \sum_i w_i - \sum_{i \in \ell} w_i \right ]}
\end{equation}

respectively. With the null hypothesis that there is no eclipse, the $n$-$\sigma$ upper limit to the flux density is given by $n\sqrt{\mathrm{Var}(\delta)}$.

The depth of transit, associated uncertainty and the 3-$\sigma$ upper limit to the flux density for each of the five lightcurves of figure \ref{fig:lc} are displayed in table \ref{tab:limits}.

\begin{table*}
\centering
\caption{Measured eclipse depths and upper limits to the radio flux density from HD 189733 b in several parts of the bandwidth.}
\label{tab:limits}
\begin{tabular}{lccccc}
\hline
Bandpass & Frequency & Eclipse & Standard deviation  & Significance & 3-$\sigma$ upper limit to \\
& range (MHz) & depth (mJy)& of depth (mJy) & & flux density (mJy) \\
\hline
all & 307.0 - 347.0 & -21.6 & 27.1 & -0.8 $\sigma$ & 81 \\
(i) & 308.6 - 312.7 & +15.6 & 31.4 & +0.5 $\sigma$ & 94 \\
(ii) & 319.3 - 325.6 & +6.8 & 26.5 & +0.3 $\sigma$ & 80 \\
(iii) & 329.8 - 332.7 & -50.3 & 33.9 & -1.5 $\sigma$ & 102 \\
(iv) & 335.2 - 339.3 & -56.5 & 28.0 & -2.0 $\sigma$ & 84 \\
\hline
\end{tabular}

\medskip
\end{table*}

\begin{figure}
\includegraphics[angle=270,width=8.25cm]{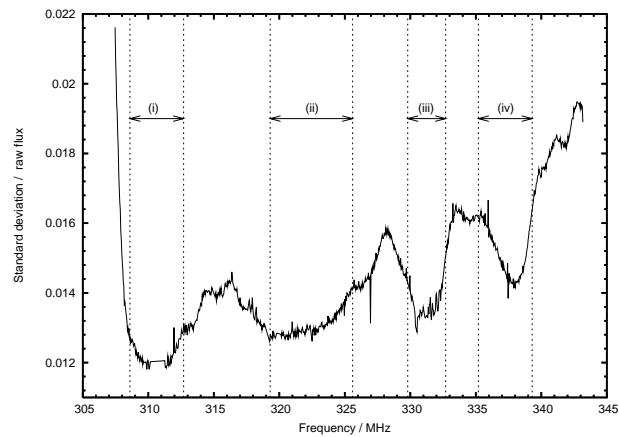}
\caption{Standard deviation of each channel of raw ON data, normalised by the mean flux of each channel as a function of frequency. The four regions indicated are those parts of the bandwidth for which we generate lightcurves (see text).
}
\protect\label{fig:sdflux}
\end{figure}

\begin{figure*}
\subfigure{\label{fig:lc_all}
(a)
\includegraphics[angle=270,width=8cm]{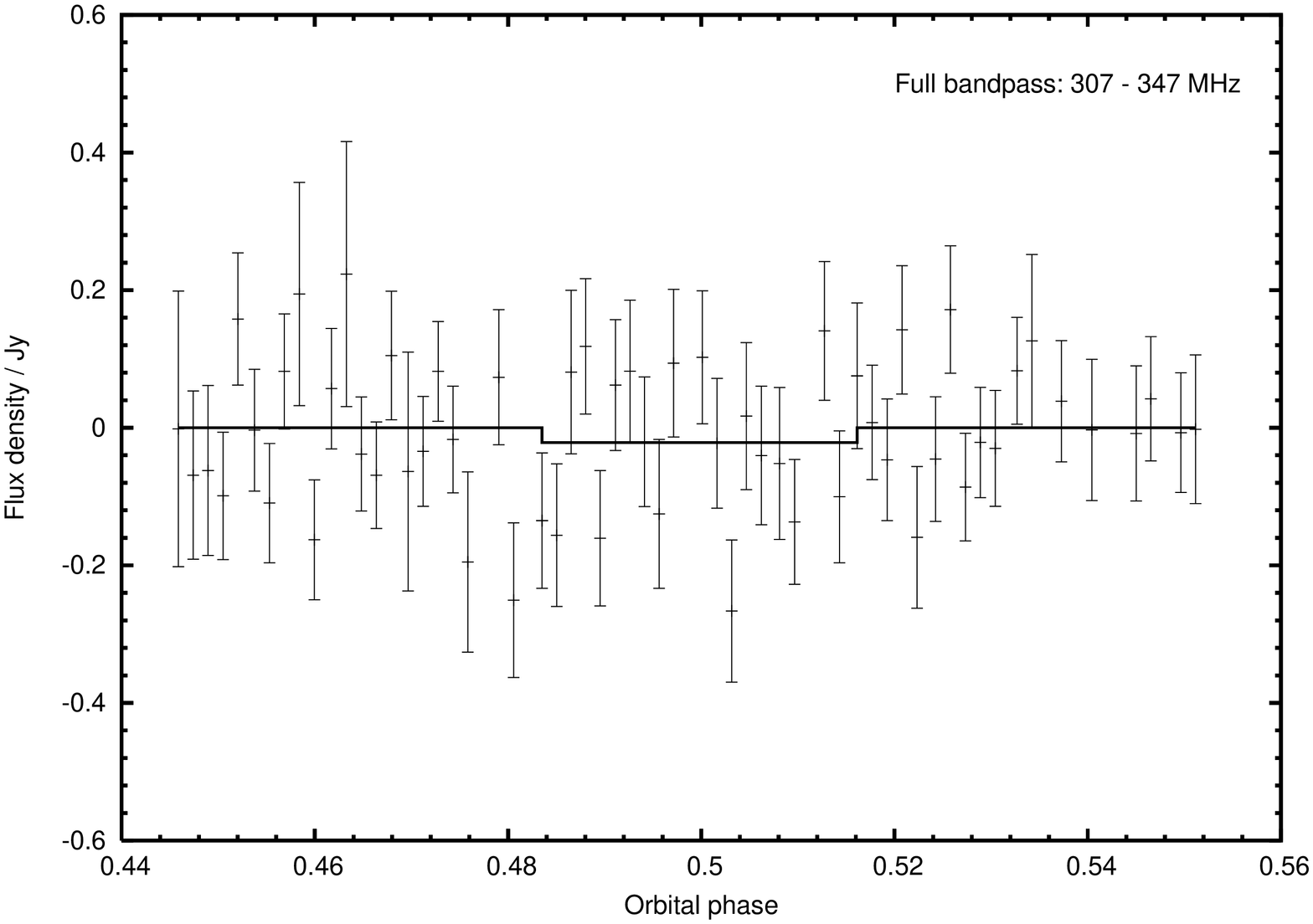}
}\\
(b)
\subfigure{\label{fig:lc_1}
\includegraphics[angle=270,width=8cm]{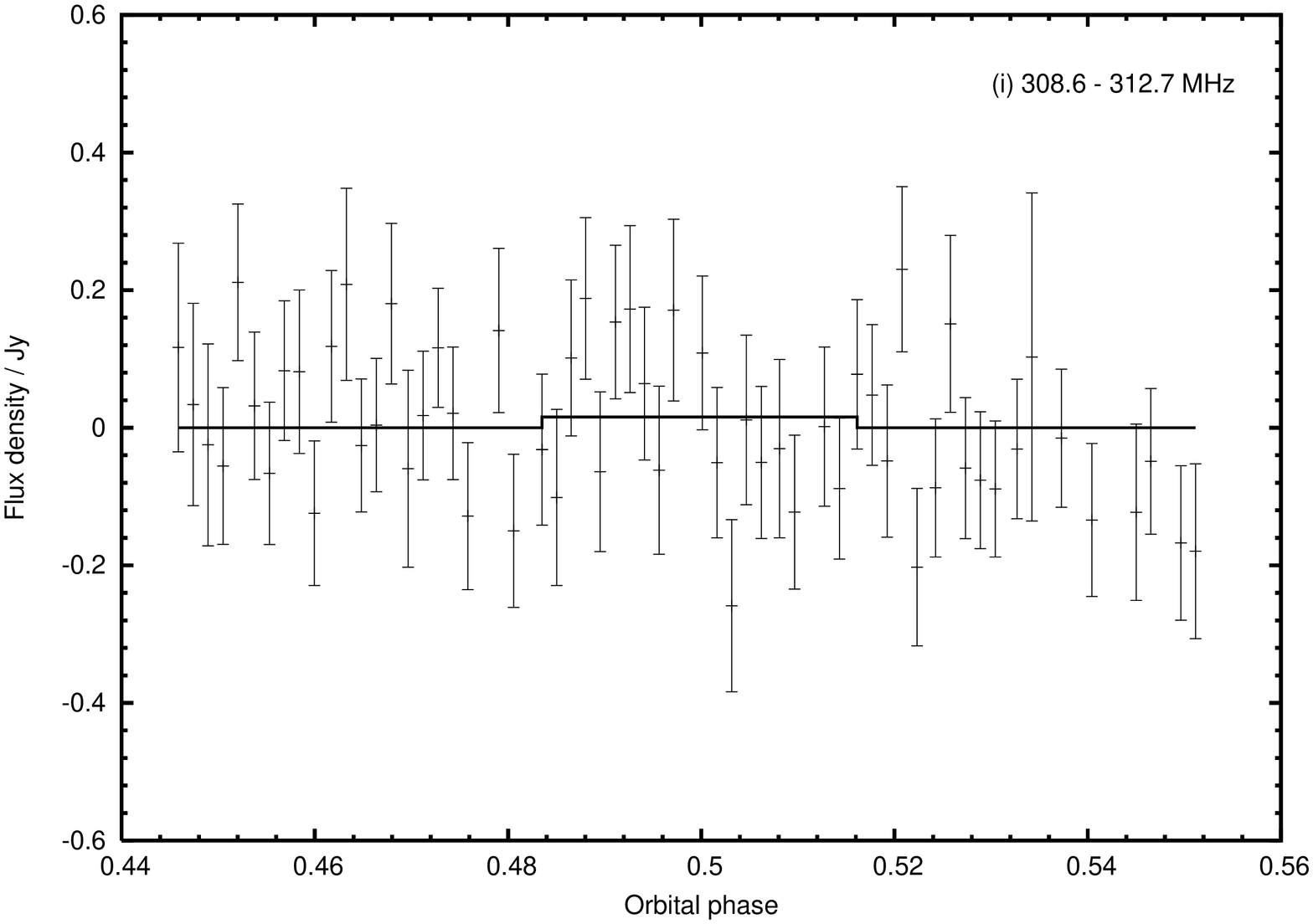}
}
(c)
\subfigure{\label{fig:lc_2}
\includegraphics[angle=270,width=8cm]{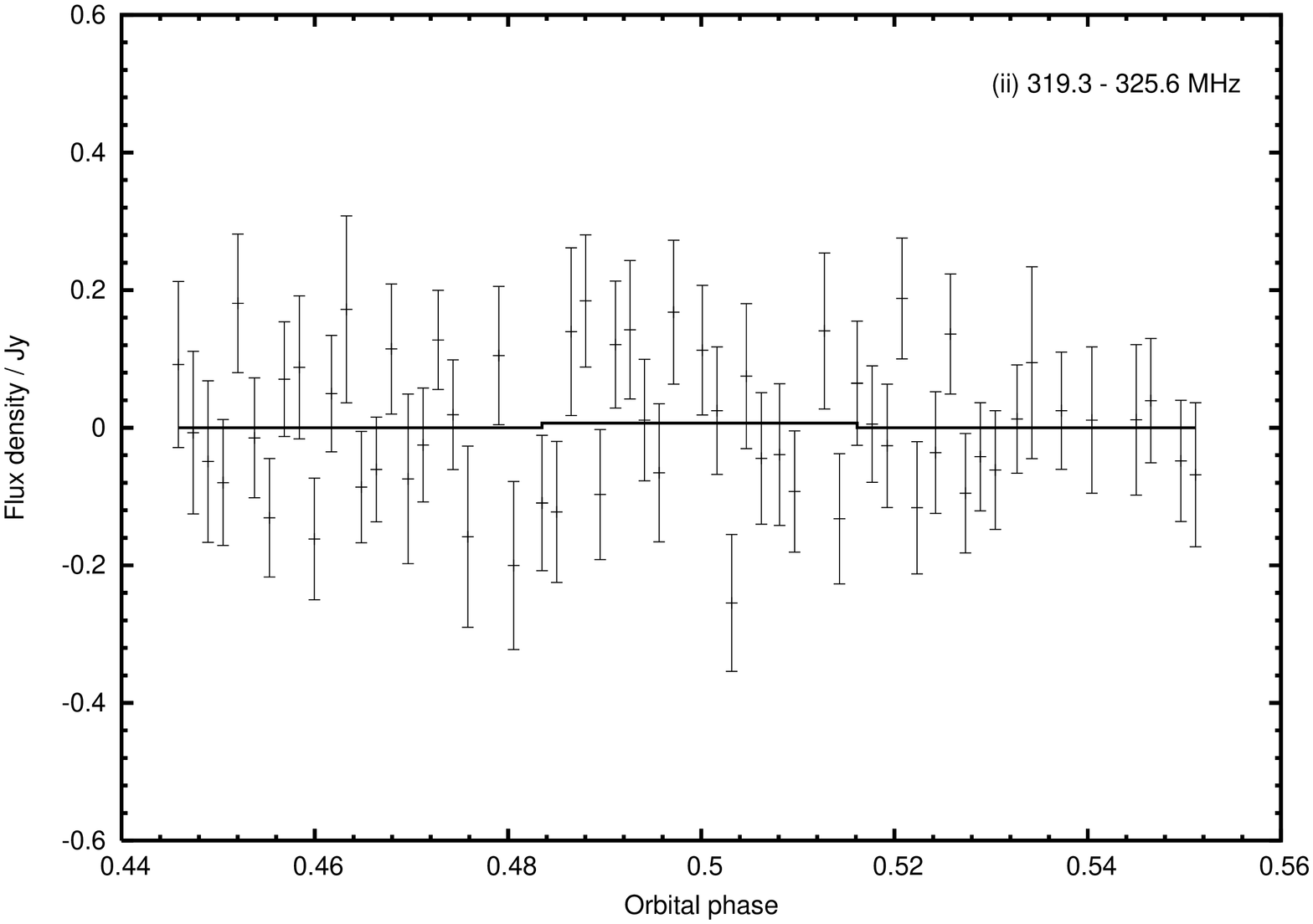}
}
(d)
\subfigure{\label{fig:lc_3}
\includegraphics[angle=270,width=8cm]{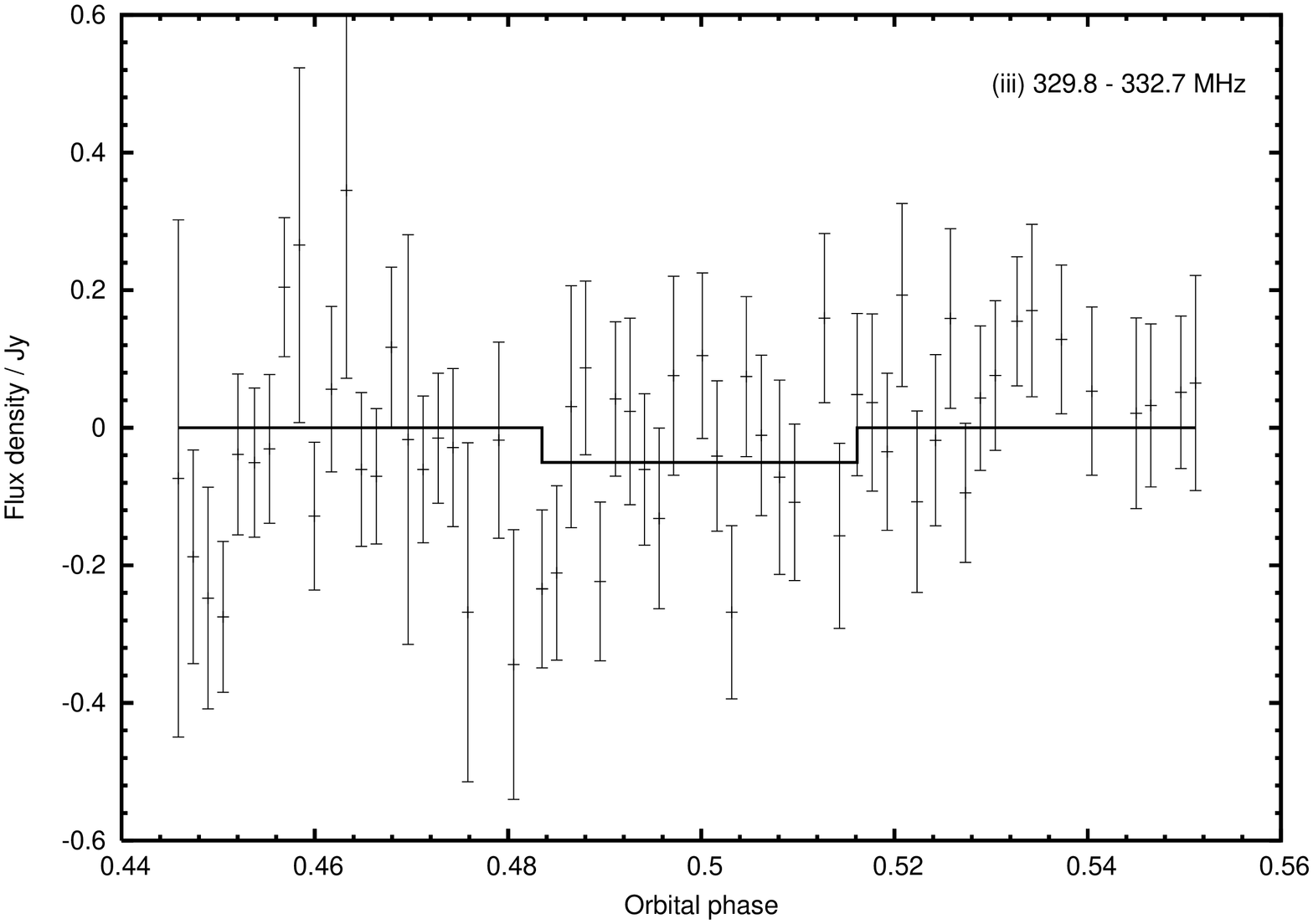}
}
(e)
\subfigure{\label{fig:lc_4}
\includegraphics[angle=270,width=8cm]{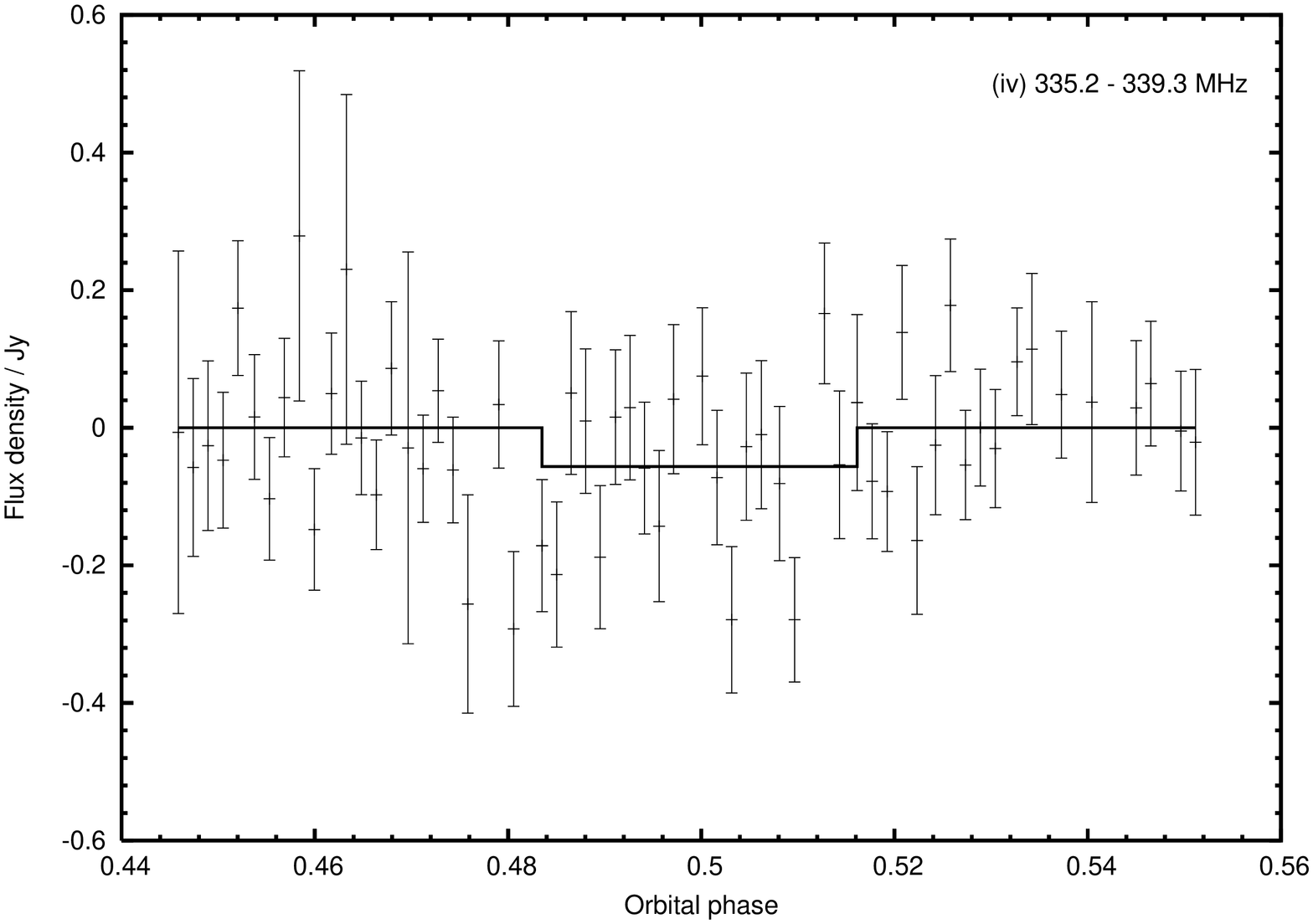}
}
\caption{Calibrated lightcurves of HD 189733 in different frequency ranges, (a) the whole bandpass, and (b), (c), (d) and (e), the four bandpasses (i), (ii), (iii) and (iv) respectively (see figure \ref{fig:sdflux}). Also plotted in each panel is the best fit to the data (see text for details).
}
\protect\label{fig:lc}
\end{figure*}

Over the whole bandwidth, we do not detect a significant eclipse, but we are able to place a 3-$\sigma$ upper limit of 81 mJy to the flux density from the planet. Of the four selected subsets of the bandpass, we find insignificant brightening (consistent with a constant flux model) at the predicted time of secondary eclipse in regions (i) and (ii). Eclipses with significances of 1.5-$\sigma$ and 2.0-$\sigma$ are found in regions (iii) and (iv), respectively.

It is possible that these 'detections' of the radio eclipse are caused by scatter in the lightcurve, and are not real; we do not understand why it should be any easier to detect the eclipse signal in these particular narrow frequency ranges than in any other. We are unable to conclude with any confidence that we have detected the eclipse, although these offer the possibility that we are tantalisingly close to detecting the first radio emission from an extra-solar planet.

\section{Discussion}
\label{sec:discuss}

\subsection{Interpretation of our results}

There are three aspects of planetary cyclotron maser emission that might lead to a non-detection: (i) the planet does not radiate at the observed frequencies, (ii) any emission is insufficiently strong for us to detect and (iii) the emission is beamed out of the line of sight

We consider first the case where the magnetic field strength of the planet does not allow radiation to be generated at the observed frequency range. It seems unlikely that the magnetic field is too strong, since it is expected that the low-frequency tail of the spectrum will be fairly flat \citep{Gmr05}. The fact that our higher frequency observations produce more encouraging results, however, hints at the possibility that the magnetic field is too strong to produce emission at the low-frequency end of the observed bandwidth.

Jupiter's cyclotron maser emission is observed from about 0.3 to 40 MHz, corresponding to a highest local value of the magnetic field strength of about 14 G. In order for us to detect emission at 307 - 347 MHz, the emission would have to be generated in a region where the local field strength is $\sim$110 G. Since it is probable that Jupiter's emission originates well above the cloud deck, it is possible that emission from an extra-solar planet could be produced closer to the planetary surface, where the local field strength would be greater. However, if no emission is generated in regions where the magnetic field strength is about 110 G or stronger, all of the emission would occur at frequencies below our observing bandwidth and no signal would be detected.

The second case is that in which the planet may produce emission in the observed frequency range, but at an insufficient intensity to be detected by our observations. According to the theoretical model of \cite{Jardine}, a stellar coronal density between 2.0 and 2.5 times greater than the solar value implies the existence of a population of electrons sufficient to generate radiation that would be detected at our 3-$\sigma$ confidence level. Therefore, the stellar coronal density must be at least twice solar in order for us to have been able to detect the emission, even if our observations were conducted at the optimum frequency.

Third, the emission may be beamed in a cone, in a manner similar to Jupiter's Io-controlled radiation. If this is the case it may be that the beam does not cross our line of sight and we would not observe any radio emission. Predicting this possible beaming effect is beyond the scope of our current model, however.

Additionally, there exist potential observational reasons for the lack of detection. One possibility is that the integration time of individual spectra was too long, at 1 s, given that Jupiter's emission displays variability on very short timescales. The emission from Jupiter that varies on the shortest (millisecond) timescales, however, is that controlled by the moon Io, whereas the emission not associated with Io varies on timescales of order 1s or greater \citep{Carr83}. Furthermore, even if there is short period variability of the emission from HD 189733b, we should still be able to observe the mean level of this flux drop as the planet goes into eclipse.

Another possibility is that the eclipse was masked by the variability of another radio source in either the ON or the OFF beam. This is discussed in Section \ref{sec:obs}, where we suggest that it is unlikely that the eclipse signal is mimicked or masked by the variability of other objects, given the nature and timing of the expected signal.
}

\subsection{Comparison with previous observations}

The upper limits calculated in this work compare favourably with those derived in previous work on different systems. \cite{Bastian00} observed seven extra-solar planetary systems with the Very Large Array (VLA), with typical 1 $\sigma$ sensitivities of around 50 mJy at 74 MHz and 1-10 mJy at 333 MHz. More recently, \cite{Lazio04} observed 5 systems with the VLA and derived 2.5 $\sigma$ upper limits of 218 to 325 mJy. A further limit of 150 - 300 mJy using the VLA was placed on the radio flux from the extra-solar planet $\tau$ Boo \citep{Lazio&Farrell07}. Tighter (2.5-$\sigma$) upper limits of 7.9 and 15.5 mJy were placed on the emission from $\epsilon$ Eri b and HD 128311b respectively by \cite{George&Stevens07} at 150 MHz using the Giant Metrewave Radio Telescope. Ours are the first eclipse observations, and they produce upper limits consistent with previous observations.

\section{Conclusions and future prospects}

We have determined new upper limits to the radio flux density from the transiting planet HD 189733 b. Our 3-$\sigma$ limit to the flux density over 307 - 347 MHz is 81 mJy. Of the four subsets of this bandpass selected for their noise properties, we detect no eclipse in two of them, but we detect marginal eclipses in the other two, the strongest of which is a 2-$\sigma$ detection. Further observations at the time of secondary eclipse in the future may enable us to improve the signal-to-noise further in order to determine whether the signal apparently detected is real or not.

Ultimately, it may be that a more sensitive, low frequency instrument such as the forthcoming Low Frequency Array (LOFAR) is required to make the first detection of radio emission from the magnetosphere of an extra-solar planet. Not only will LOFAR be more sensitive than existing instruments, but it will be able to observe at frequencies of 10 - 240 MHz \citep {Farrell04}, enabling the detection of emission from planets with weaker magnetic fields than at present.

\section{Acknowledgements}

AMSS acknowledges the financial support of a UK PPARC / STFC studentship. This research has made use of the VizieR catalogue access tool, CDS, Strasbourg, France. The authors wish to thank the anonymous referee, whose helpful comments led to improvements in the manuscript; and Keith Horne, for useful discussions.

\bibliographystyle{mn2e}
\bibliography{iau_journals,review}
\bsp

\end{document}